\documentclass[12pt,preprint]{aastex}

\shorttitle{Protostellar Envelopes} 
\shortauthors{Chiang et al.}

\begin{document}

\title{Constraining the Earliest Circumstellar Disks and their Envelopes} 
\author{Hsin-Fang Chiang\altaffilmark{1}, Leslie W. Looney\altaffilmark{1}, 
Konstantinos Tassis\altaffilmark{2}, Lee G. Mundy\altaffilmark{3}, 
Telemachos Ch. Mouschovias\altaffilmark{4}}

\altaffiltext{1}{Department of Astronomy, 
University of Illinois at Urbana-Champaign, 1002 West Green Street, Urbana, IL 61801; 
hchiang2@uiuc.edu}
\altaffiltext{2}{Department of Astronomy and Astrophysics and 
the Kavli Institute for Cosmological Physics, 
University of Chicago, Chicago, IL 60637}
\altaffiltext{3}{Department of Astronomy, 
University of Maryland, College Park, MD 20742}
\altaffiltext{4}{Departments of Physics and Astronomy, 
University of Illinois at Urbana-Champaign, 1002 West Green Street, Urbana, IL 61801}

\begin{abstract}
Using interferometric data from BIMA observations, combined 
with detailed modeling in Fourier space of the physical structures 
predicted by models, we constrain the circumstellar envelope parameters 
for four Class 0 young stellar objects, 
as well as their embedded circumstellar disks. 
The envelopes of these objects are still undergoing collapse,  
and theoretical collapse models can be 
compared to the observations.  
Since it has been suggested in a previous study that both the Larson-Penston and Shu 
similarity solutions underestimate the age of the system, we adopt 
Tassis \& Mouschovias' model of the collapse process, 
which includes all relevant magnetic fields effects.  
The results of the model fitting show a good consistency 
between theory and data; 
furthermore, no age problem exists since the Tassis \& Mouschovias' model 
is age independent for the first 255 kyr. 
Although the majority of the continuum dust 
emission arises from the circumstellar envelopes, these objects 
have well known outflows, which suggest the presence of circumstellar disks.  
At the highest resolution, most of the large-scale envelope emission is 
resolved out by interferometry, but the small-scale residual emission remains, 
making it difficult to observe only the compact disk component.  
By modeling the emission of the envelope and subtracting it from the total emission, 
we constrain the disk masses in our four systems to be comparable to or smaller than 
the typical disk masses for T Tauri systems. 
\end{abstract}

\maketitle

\section{Introduction}
The standard scenario of low-mass star formation starts at the collapse of 
prestellar cores and the formation of central protostellar objects.
These young stellar objects (YSOs) evolve through the so-called
Class 0, I, II, and III stages, which are thought to be a temporal sequence
\citep[e.g.,][]{Lada1984,Adams1987,Andre1993,Andre2000PPIV}. 
In the earliest stage, i.e. Class 0 stage, 
when the central YSO is just forming inside the surrounding envelope
\citep[of mass $\approx$ a few solar masses, e.g.,][hereafter LMW2000]{LMW2000},  
the envelope is still undergoing gravitational collapse 
onto the circumstellar disk.  
The YSO powers the bipolar outflows, which 
carve away the polar region of the
envelope by entraining envelope material 
and widening their opening angles 
\citep[e.g.,][]{Bachiller1996ar,Arce2006,Seale2008}.
At this early time, the envelope mass is $>$85\% of the
system mass \citep[][hereafter LMW2003]{LMW2003}.
As the system evolves, the envelope loses mass as material
is transported down through the circumstellar
disk onto the protostars or carried away with the outflows.
Eventually, the YSO circumstellar structure is dominated
by the disk (a hundredth of a solar mass, e.g., Andrews \& Williams 2005).
The circumstellar disk evolves, 
presumably becoming a planetary system like the Solar System.   

The initial collapse process of low-mass protostars is often described
by self-similar isothermal solutions, which are a continuum of solutions
\citep[e.g.,][]{Whitworth1985} 
that range from the ``inside-out'' collapse solution  
\cite[][hereafter the Shu solution]{Shu1977} 
to the Larson-Penston solution
\citep[][hereafter the LP solution]{Larson1969,Penston1969,Hunter1977}.
These models generally
obtain an inner core with a power law density profile $\rho \propto r^{-3/2}$ that increases in radius with time, 
surrounded by a $\rho \propto r^{-2}$ envelope.  
The theoretical density profiles from these solutions 
have been compared to observations 
of the dust continuum emission 
\citep[e.g., LMW2003;][]{Harvey2003a,Jorgensen2005}, 
but the models cannot fit the observations  
with reasonable physical parameters 
(required age is too low; see LMW2003), 
which consequently hints at the need for 
more sophisticated theoretical models that include more of the
essential physical processes of the collapse mechanisms, for example, 
turbulence and/or magnetic fields.

The theory of turbulence-induced star formation postulates that
turbulence causes over-densities 
and is thus responsible for the core formation in molecular
clouds, while magnetic fields are not dynamically important and do not
have a significant impact on this process   
\citep[see reviews of ][]{MacLow2004,Elmegreen2004}.
At this time, there are no predictions of the density
of a protostellar object that is produced by turbulence-induced
collapse. Moreover, 
the non-thermal contribution of the observed linewidths is small 
in evolved, collapsing molecular cloud cores  
\citep[e.g.,][]{bm89, bg98, kjt07}.
Whether turbulence plays an important role 
in the formation and evolution of protostellar fragments (or cores) 
is still under debate. 
On the other hand, the theory of ambipolar-diffusion-initiated star
formation predicts the formation of thermally and magnetically 
supercritical protostellar cores inside magnetically subcritical parent clouds
\citep[see reviews of][]{Mouschovias1996, Mouschovias1999}.
In the framework of the ambipolar-diffusion-induced collapse, there
are extensive studies of the dynamics of the 
prestellar phase \citep[e.g.,][]{tm07a,tm07b,tm07c} and the accretion process after a
protostar has formed at the center of the core 
\citep[][hereafter TM2005]{Tassis2005a,Tassis2005b}, which can be compared to observations.

TM2005 have constructed a six-fluid MHD simulation to study the 
accretion process of matter from a molecular cloud core onto 
a protostellar object in the presence of magnetic fields.
In their model, they track the evolution of magnetic flux and six kinds of 
particles (neutral molecules, atomic and molecular ions, electrons, neutral 
grains, negatively-charged grains, and positively-charged grains, 
among which only the electrons are assumed to be attached to the magnetic field lines)  
in a self-gravitating, accreting molecular cloud core.  
The simulation starts with a magnetically supported parent cloud.   
Ambipolar diffusion leads to the formation of a thermally and magnetically supercritical core 
that begins to contract dynamically.  Its innermost part reaches nearly 
hydrostatic equilibrium while its outer part still undergoes infall.  
At the moment when the hydrostatic
protostellar core has just formed at the center (called {\it t}~=~0), 
the inner core region including the protostar 
is replaced by a central sink to facilitate the calculation.  
As time progresses and mass and magnetic flux 
accrete onto the protostar from the envelope, a region of enhanced
magnetic field, called the ``magnetic wall'', forms and drives 
an outward-propagating shock. 
Behind the shock, gravity dominates over the magnetic forces 
and reaccelerates the neutral particles, which continue to accrete
onto the protostar until the next magnetic wall is formed. 
The magnetic wall forms and disperses 
in a quasi-periodic manner.  
Because of the presence of the magnetically controlled bursts, accretion from 
the envelope onto the protostar is episodic over the first 255 kyr.

Interferometric data of Class 0 objects provide the best means to test
these theories.
One of the features of an interferometer is 
the ability to spatially filter
emission.  Indeed, dust continuum observations
of young protostars have been often used to peer inside the bright
envelope to reveal the young,
compact circumstellar disk 
\citep[e.g.,][]{Keene1990}.
Dust continuum emission is often used, but molecular lines can also be
excellent tracers of specific conditions.  
However, using molecular lines to trace the disks do have some difficulties 
in the case of the youngest stars: 
(1) active accretion and outflow processes at multiple scales,
(2) chemistry effects and evolution,
and (3) shocks in the outflow, the disk, and the disk/envelope interface region. 
All of these contaminate the preferred disk-only tracers and make it
difficult to disentangle any molecular result without
a good understanding of the envelope structure 
derived from the dust continuum modeling \citep[e.g.,][]{Brinch2007}.

Regardless of the tracers used, more intricate 
theoretical models than the ``inside-out'' collapse can be tested observationally. 
In this paper, we build on the observational data of LMW2000: 
$\lambda$~=~2.7~mm 
dust continuum images of 24 young stellar sources 
with sensitivity to spatial scales
from 0$\farcs$5 to 50$\arcsec$.  
A discussion of the data acquisition and images can be found in that paper.  
We follow the work of LMW2003, which presented
modeling of the envelope emission of Class 0 objects, 
and use the 4 sources from that work with the highest signal-to-noise ratio
(NGC 1333 IRAS 4A, NGC 1333 IRAS 4B, NGC 1333 IRAS 2A, and L1448 IRS 3B).
We compare the predicted density profiles from TM2005 to these data and
comment on the results with respect to those found in LMW2003.

\section{Sources}
All 4 sources modeled in this study are in the Perseus molecular cloud, 
a low mass star forming region probably in the vicinity of 
the massive star forming region Per OB2 association.  
Dense cores and YSO candidates 
at all evolutionary stages 
(Class 0, I, II, and III) have been found in Perseus 
via radio and infrared observations. 
IC 348 and NGC 1333 are the two main dense clusters and 
other smaller groups like L1448, L1455, Barnard 1, and Barnard 5 
are also associated with many low mass protostars  
\citep[e.g.][]{Enoch2006, Jorgensen2006}. 
The exact distance to the Perseus molecular cloud 
is still uncertain 
and ranges from 220 to 350 pc.  
The smaller value is based on the distance-interstellar extinction relation 
using photometry \citep{Cernis1990};  
the distance may be the larger value 
if the Perseus molecular cloud is physically related to 
the Per OB2 association, Hipparcos parallax distance of 318 $\pm$ 27 pc
\citep{deZeeuw1999}. 
Since Perseus is composed of a long chain of dense clouds 
with a total length of about 30 pc, there may be a distance gradient  
or it may be composed of several layers of clouds.  
\citet{Cernis1993}
showed a distance difference from the eastern part ($\sim$260 pc) to 
the western part ($\sim$220 pc).
In this study, we adopt a distance of 350 pc for NGC 1333 and 300 pc for L1448 
as in LMW2003 to facilitate comparison.    
Since we are using roughly the upper limit of the distance, 
this assumption may lead to overestimates of the envelope 
and disk masses and underestimates of the source size.   

\subsection{NGC 1333 IRAS 4 }
NGC 1333 is a reflection nebula with 
mainly two embedded protostellar clusters 
in the L1450 dark nebula in the Perseus molecular cloud.   
The age of the young cluster is about 1 Myr, 
estimated by the fraction of infrared excess sources and 
a K-band luminosity function comparison \citep{Lada1996},    
consistent with the age estimated by brown dwarf studies  
\citep[e.g.][]{Wilking2004}.  
The plentiful jet and outflow activities driven by YSOs in this region 
also imply that it is an  
active star forming region at an early stage of evolution 
\citep[e.g.,][]{Bally1996,Knee2000}. 
 
The multiple system NGC 1333 IRAS 4 (hereafter IRAS 4) 
contains mainly three groups of sources 
designated as 4A, 4B, and 4C 
\citep[][]{Sandell2001}. 
The brightest Class 0 component IRAS 4A 
has been detected as a binary system separated by 1.8$\arcsec$ 
with a common circumbinary envelope  
(\citealt{Lay1995}; LMW2000)  
A highly collimated N-S molecular outflow driven by IRAS 4A2 has been observed 
with estimated dynamical age of about 6000 yr  
\citep[][]{Blake1995,Choi2005}. 
A dimmer southern outflow probably driven by 4A1  
has been mapped in HCN and SiO,   
but no northern counterpart has been detected 
\citep{Choi2001,Choi2005}.  
SMA polarimetric observations have shown the magnetic field 
geometry of IRAS 4A and supported the 
magnetic theory of star formation   
\citep{Girart2006}. 

IRAS 4B is a binary system
with a separation of $\sim$10$\arcsec$ between 4BW and 4BE \citep{Sandell2001};
4BE is also named as 4B', 4BII, or 4C in different references. 
Note that 4C is mostly used as the name of another millimeter object 
$\sim$50$\arcsec$ northeast of 4A. 
A compact collimated outflow driven by IRAS 4BW has been observed 
\citep[e.g.,][]{Choi2001,DiFrancesco2001}  
and shown a short dynamical timescale. 

\subsection{NGC 1333 IRAS 2A}
NGC 1333 IRAS 2 (hereafter IRAS 2) 
has been resolved into 3 sources \citep{Sandell2001}   
including two Class 0 protostars, IRAS 2A and IRAS 2B, 
and a starless core IRAS 2C  
\citep{Jorgensen2004_IRAS2}. 
Two CO outflows have been mapped: 
one in the NNE-SSW direction and the other in the E-W direction   
\citep[e.g.,][]{Knee2000}.  
Since these two outflows are orthogonal to each other and 
have quite different properties, they may have different driving sources.  
The E-W outflow may be driven by IRAS 2A, 
while the NNE-SSW outflow may be driven by IRAS 2C  
\citep{Knee2000}.  
It's also possible that both are driven by IRAS 2A 
which may be an unresolved binary.  
IRAS 2B may drive the third outflow in this region but 
only the blue-shifted lobe has been identified  
\citep{Knee2000}.  

\subsection{L1448 IRS 3B}
The star forming region L1448 is located $\sim 1\,^{\circ}$ southwest 
of NGC 1333 in Perseus and 
contains many YSOs:  
Class I source L1448 IRS 1, 
Class 0 sources L1448 IRS 2, L1448 IRS 3 (also known as L1448 N), 
and L1448-mm (also know as L1448 C)  
are the most well-known  
\citep[e.g.][]{Jorgensen2006, OLinger2006}. 
Many of these YSOs have been resolved into binary systems;  
for example, L1448 IRS 3 and L1448-mm are both binary systems.  
Multiple outflows have been found in this region.  
The huge molecular outflows emanating from L1448-mm and L1448 IRS 2 
are almost parallel to each other, 
and some of them even have multiple overlapping components 
\citep[e.g.][]{Wolf-Chase2000,Kwon2006,Tobin2007}.  

L1448 IRS 3 is 
composed of 3 sources, 
among which IRS 3A (L1448 N:A) and IRS 3B (L1448 N:B) 
are separated by 7$\arcsec$ and 
have a common envelope in a protobinary system, 
and IRS 3C (L1448 NW) is 20$\arcsec$ northwest of them (LMW2000).   
All of them are Class 0 objects, 
except that IRS 3A is slightly closer to  
the transition between Class 0 and Class I 
\citep{OLinger2006}. 
Two interacting outflows driven by IRS 3A and IRS 3B, respectively, 
have been studied by \citet{Kwon2006}.   

\section{Modeling}
To compare the theoretical models of TM2005 and the LMW2000 observations,
we characterize the observed dust continuum emission, which depends on the
dust density, grain properties, and temperature. 

\subsection{Density Profiles}
In the theoretical TM2005 model, the physical parameters of the envelope are 
shown to have a quasi-periodic variation beginning at the time of 
about 15 kyr after the formation of a hydrostatic protostellar core. 
The density goes through a cycle profile that is 
largely invariant with time 
from 15 to 255 kyr.
This implies that, unlike the LP or Shu models, we cannot estimate the age
of the source based on its density profile alone.

\begin{figure}
\includegraphics[angle=270,width=1.0\textwidth]{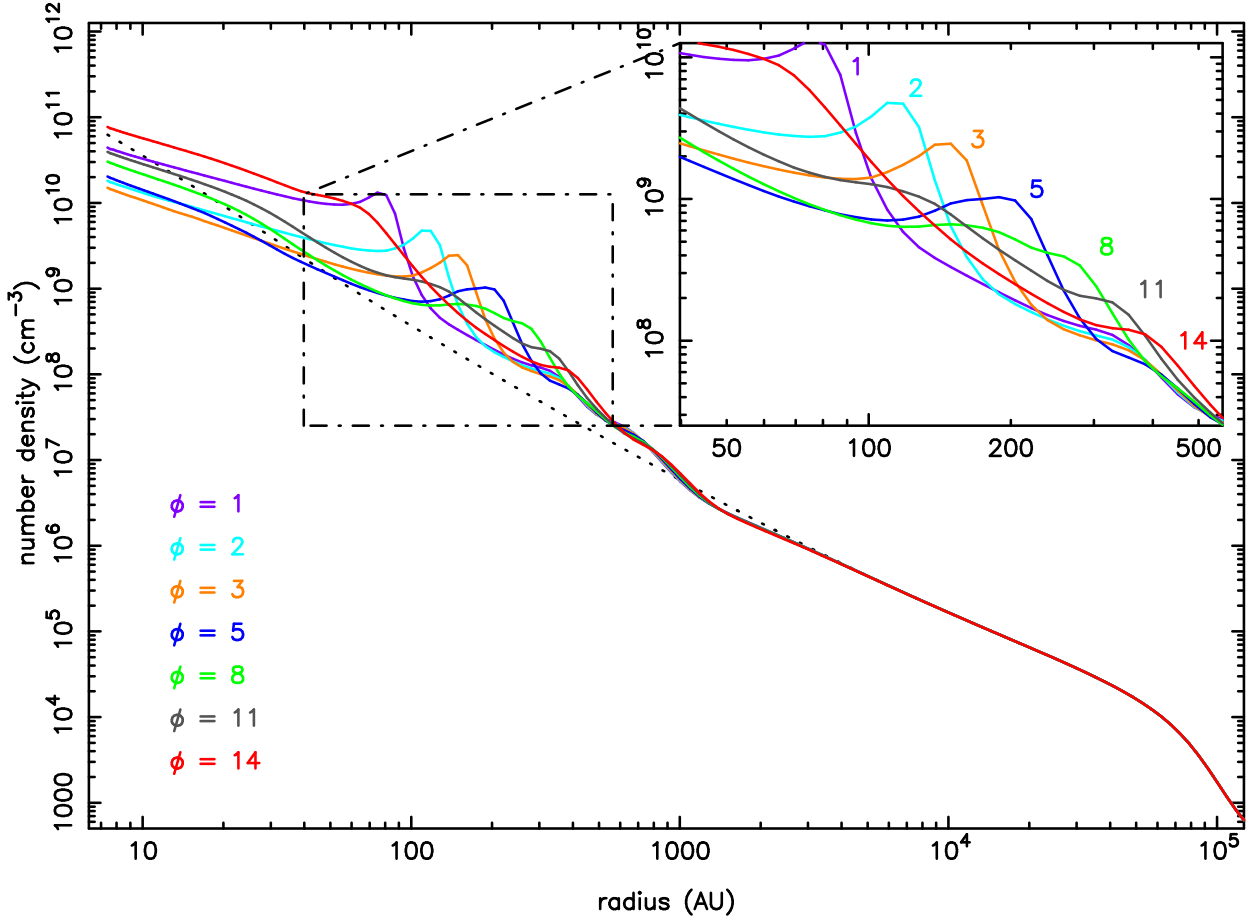}
\caption{The number density profiles of neutral particles 
in a typical magnetic cycle at different phases from the TM2005 model. 
These curves show phase 1, 2, 3, 5, 8, 11, and 14, which correspond to 
{\it t}~=~33750, 34000, 34250, 34750, 35500, 36250, and 37000  
yr after formation of the central protostar. 
The shock driven by the ``magnetic wall'' forms, propagates outward, and 
disperses throughout a cycle.  
Although this is a particular cycle, 
no obvious differences are found between cycles, that is, 
every magnetic cycle goes through similar phases, 
except that the period of the cycle varies as the system evolves  
(see TM2005a,b for more details). 
}
\label{figDenProf}
\end{figure}

In this study, we adopt a typical set of density profiles,
averaged along the characteristic scale height, from one of the 
magnetic cycles predicted by the TM2005 model. 
The density structure repeatedly goes through phase $\phi =$ 1 to 15   
in an evolution cycle, as shown by the solid curves 
in Fig. \ref{figDenProf}.  
Here we only plot the more representative phases.  
The dotted curve in Fig. \ref{figDenProf} shows the initial density profile
at the time hydrostatic equilibrium is established in the central region
of the core ({\it t}~=~0),
which nearly follows a power-law relation of index $-1.7$.  
We can choose density profiles of any cycle for this study,
as they are all similar. 
In other words, different cycles have the same predicted density profiles 
and quantitatively similar phases.  
The age of the chosen magnetic cycle is from {\it t}~= 33,750 to 37,250 yr 
after the formation of the central hydrostatic core.
By this time, the series of magnetic cycles has been well established. 

The model cloud in TM2005 has an equatorial radius of 4.23 pc. However,
the thermally and magnetically supercritical fragment extends to approximately 9000
AU and contains $\sim$9 M$_\odot$. Outside this region, hardly any evolution 
takes place (see Fig. \ref{figDenProf}). In order to compare with observations, we
truncate the density profiles at an outer radius within which the dynamical
infall takes place. This is the "envelope", at the center of which the
protostar is forming. By scaling the density profiles, the theoretical TM2005
model is applied to observed objects of different masses. (This kind of
scaling of the density profiles implies a corresponding scaling of the magnetic 
field profiles, but, since no information on the magnetic field is
available from observations, we do not discuss this implication further in this
paper.)

Although the TM2005 model has a flattened morphology \citep{fiedler1993},
we only use a spherical density profile.
The flattened molecular cloud has a radius of several pc,
which is much larger than the scale studied here.
LMW2003 has tested geometric effects by elongating the envelope,
and shown that flattening the envelope will artificially make a
steeper flux density profile in  {\it u-v} space.
However, the analysis normalized the maximum flux density, which is different
than the effect studied here.
To better compare the effect a flattened envelope has on our modeling, we
modified our spherical envelope model by multiplying the predicted image
with a flat-shaped mask with exponentially-decayed edges similar to a
flattened envelope. The half-thickness of the ``disk'' is one-fifth of the
radius, which gives a similar flattening ratio to that in the TM2005
model.  In this case, the observed flux density is less steep at small
{\it u-v} distances, but is unchanged beyond {\it u-v} distances of 10 $k \lambda$.  
This is because fewer large-scale components, or small {\it u-v} distance
components, contribute in the flattened structure compared to a
spherical model. We fit the simulated data of a flattened envelope with
our spherical model and found that the
spherical model fits the data very well with the same density profile,
but that the mass is overestimated. In addition, the
density profile from TM2005 has been averaged along the characteristic
scale height, which also helps mitigate the effect of a flattened
envelope.  On the other hand, the vertical density profile needs to be
better modeled \citep[see the observed flattened envelope in][]{looney2007}, 
which is beyond the scope of this paper. Finally, a
circumstellar disk may exist inside the envelope and its emission is also
taken into account in the model; this is described in more detail in $\S$
3.4 and $\S$ 4 .

\subsection{Grain Properties and Temperature Profiles} 
Properties of dust grains such as composition and size 
determine their extinction and emission, as well as 
the temperature structure and 
the observable dust emission of the envelope. 
Here we adopt the grain model of \citet{WC1986}, 
in which the material mixture and size distributions 
of the Mathis-Rumpl-Nordsieck model 
\citep*[][hereafter called the MRN model]{MRN1977}
and optical constants of grains in \citet{Draine1984} are used. 
The grain model consists a mixture of uncoated graphite and silicate 
with particle sizes ranging from $0.005~\mu m$ to $0.25~\mu m$ 
and a power law size distribution of index -3.5.   
It is important to note that the
grains in TM2005 are spherical and uniform in size 
with a different chemical composition from MRN.  
However, these grain differences are not significant in the theoretical
model evolution \citep{Desch2001}.

The mass opacity coefficient $\kappa _\nu$ of the grain 
is frequency dependent and typically follows a power law 
relation $\kappa _\nu \propto \nu ^{\beta}$.   
The index $\beta$ varies with environment and is related to grain properties.  
At submillimeter wavelengths, 
the unevolved grains of the ISM have $\beta \approx 2$. 
However, in disks and dense cores  
$\beta$ decreases to $1$ 
mainly due to grain growth 
\citep[e.g.][]{Beckwith1991,Natta2007PPV}. 
Although our observations are only at a single wavelength, 
$\kappa_\nu$ of a wider range of frequency is still needed for computing 
a self-consistent temperature profile 
since radiation of all frequencies contributes to the total luminosity. 
Our model uses the MRN grain model $\beta = 2$ at optical and infrared 
wavelengths but assumes $\beta = 1$ at long wavelengths,  
and adopts $\kappa _\nu = 0.009~cm^2~g^{-1}$ at    
$\lambda$~=~2.7~mm.  
The model also assumes the dust grain properties are uniform and do not change 
with radius in the envelope.  

Temperature profiles are then considered based on the model grain properties.  
The temperature profile can be simplified as a power law with radius 
if the dust envelope is optically thin and the dust opacity 
has a power-law frequency dependence ($\kappa _\nu \propto \nu ^{\beta}$):
temperature $T \propto r^{-2/(4+\beta)}$  
assuming the central protostar is the only heating source. 
In the case of $\beta = 1$, 
the temperature is $T \propto r^{-0.4}$. 
But in our model, the inner part becomes 
optically thick as the density increases near the center (LMW2003), 
so we calculate a self-consistent temperature profile 
for each fitting using the code of  
\citet{WC1986}. 
The code takes the luminosity of the central object and 
solves the radiative transfer equation including the effects of 
both emission and extinction by the dust grains in the envelope. 
At each shell of the envelope, the luminosity is conserved.  

The self-consistent temperature profiles are important for calculating 
the emission, but they contradict
the isothermal assumptions in theoretical models. 
However, until non-isothermal theoretical models are developed, 
this is the best compromise (also see LMW2003).

\subsection{Interferometric Filtering}
Although we use the power of the interferometer to resolve out the 
large-scale features of the envelope, it is important to point out that
there is remnant envelope emission
even with high resolution configurations (e.g., LMW2000).   
There are a few reasons for this.  First, the envelope emission 
is power-law like (e.g., LMW2003), so the expected interferometric response   
from a Gaussian \citep{Wilner1994} is not applicable.  
Second, the inner edge cutoff is abrupt, especially with the
steep density profile.  The abruptness causes ripples in {\it u-v} space that
create power at long baselines.
The important point here is that the ability to detect a disk in the presence
of an envelope is not set by the formal noise level or Fourier components, 
but by the intrinsic ability to model the complexity
of the envelope (inner cutoff, asymmetric structure, etc.)
and the resolution of the observations compared to the disk
size and the inner cutoff of the envelope.
In other words, it is difficult to separate the youngest circumstellar disk
from the inner circumstellar envelope; revealing the embedded 
circumstellar disk still requires an understanding of the inner envelope.

Only interferometric data are used in this study.  
Single dish data can give an upper limit at zero-spacings, but should not 
be included in the fitting.  Single dish observations detect  
not only the emission from the inner envelope, but also that from the outer part of  
a cloud that is not actively involved in the protostar forming process.   
Using interferometry, only structures of size scale similar or smaller than 
the collapsing envelope are detected and modeled. 
Hence the mass and radius we are inferring in this paper are not the total mass 
or the overall size of the prestellar cores, but that of the inner envelope  
undergoing the collapse, observed at $\lambda$~=~2.7~mm.  

\subsection{Modeling Details} 
\begin{figure}  
\includegraphics[angle=270,width=1.0\textwidth]{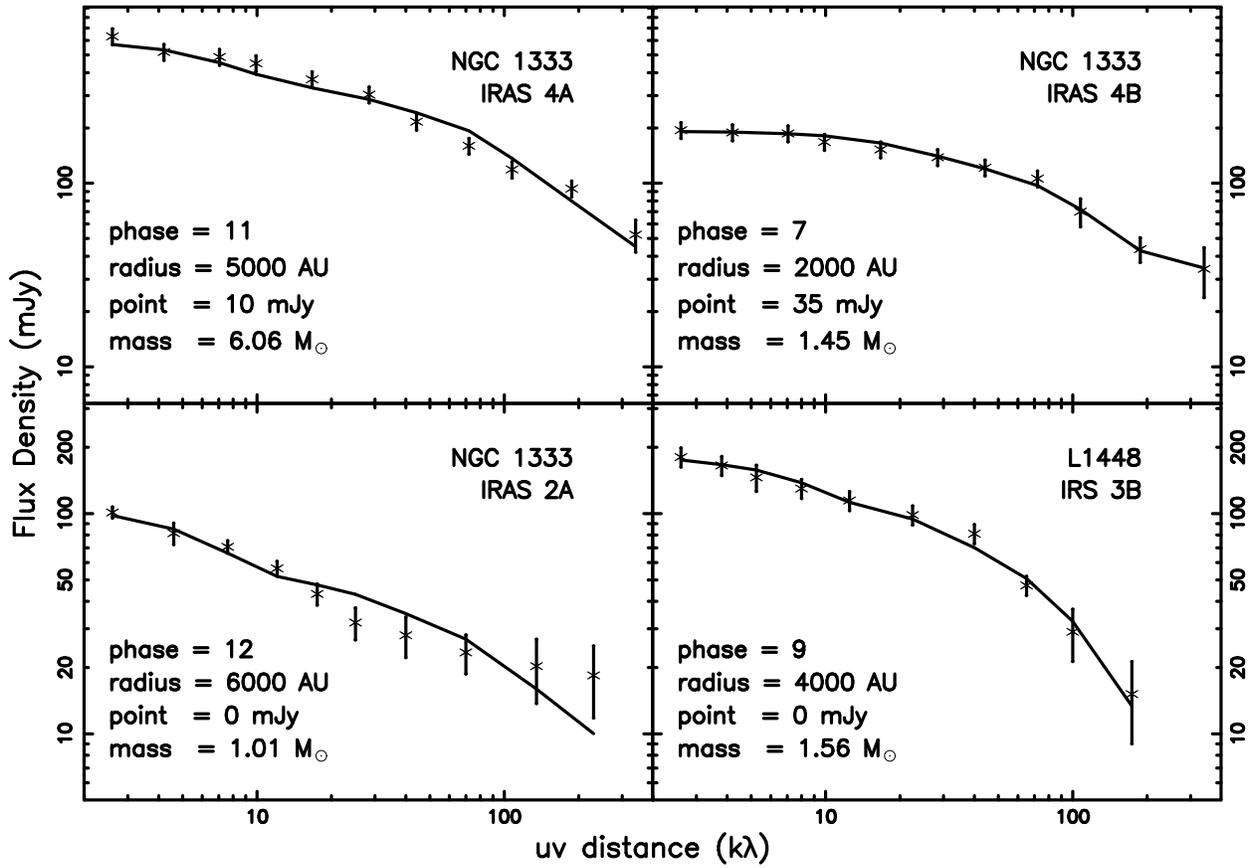}
\caption{
The flux density of the observational data and the best fit for each source  
with best fit parameters listed. 
The $\chi ^2$ values are 1.49, 0.22, 1.30, and 0.41 for 
NGC 1333 IRAS 4A, IRAS 4B, IRAS 2A, and L1448 IRS 3B, respectively. 
}
\label{figfit}
\end{figure}

The observational data we use are from LMW2000.
To compare the observations and the theoretical model, we analyze the data in
{\it u-v} space where the data are not affected by the CLEAN algorithm
or {\it u-v} sampling.
The interferometric data are binned in {\it u-v} annuli around the
source locations from LMW2000 and averaged vectorially.
The resulting {\it u-v} amplitudes for each bin are shown by asterisks 
in Fig. \ref{figfit}. 
The displayed error bars are statistical error bars
based on the standard deviation of the mean of the data points in the bin
with a minimum of 10\%, reflecting the uncertainty in the
overall calibration.
In the cases where the binary systems were separated by more
than 10$\arcsec$, the companion sources were subtracted out of the
{\it u-v} data using the large-scale images of LMW2000.
The new {\it u-v} data were remapped to confirm that the large-scale emission 
from the companion sources was not detected.
Although there may exist some residual of small-scale emission from the 
companion envelope in the {\it u-v} data, vector averaging in {\it u-v} annuli
will minimize its contribution.

The observational data shown in Fig. \ref{figfit} show the brightness
distribution plotted at various antennas, or {\it u-v}, spacings, which is the
Fourier transform of the sky brightness distribution.
In other words, power at small {\it u-v} distance represents
large-scale structures and power at large {\it u-v} distance
represents small-scale structures, i.e., a point source would be a constant at
all antenna separations.
The brightness distribution is determined by the circumstellar material surrounding
the source depending on density, temperature and grain properties.

We model the circumstellar envelope emission as arising from a
spherically symmetric dust envelope 
with TM2005 radial density and self-consistent temperature profiles and an 
embedded circumstellar disk represented by an envelope-attenuated point source.
The observed emission of most sources shows a circular symmetry and 
lack of significant internal structures
(see images of the sources in LMW2000). 
The combined radiative transfer code allows the calculation
of the expected flux as a function of radius in the image plane
or as a function of {\it u-v} distance in the Fourier plane.
This provides the best way to trace emission structure to very small
length scales, effectively the density and temperature profiles
in the inner circumstellar envelope.

\begin{figure}  
\includegraphics[angle=270,width=1.0\textwidth]{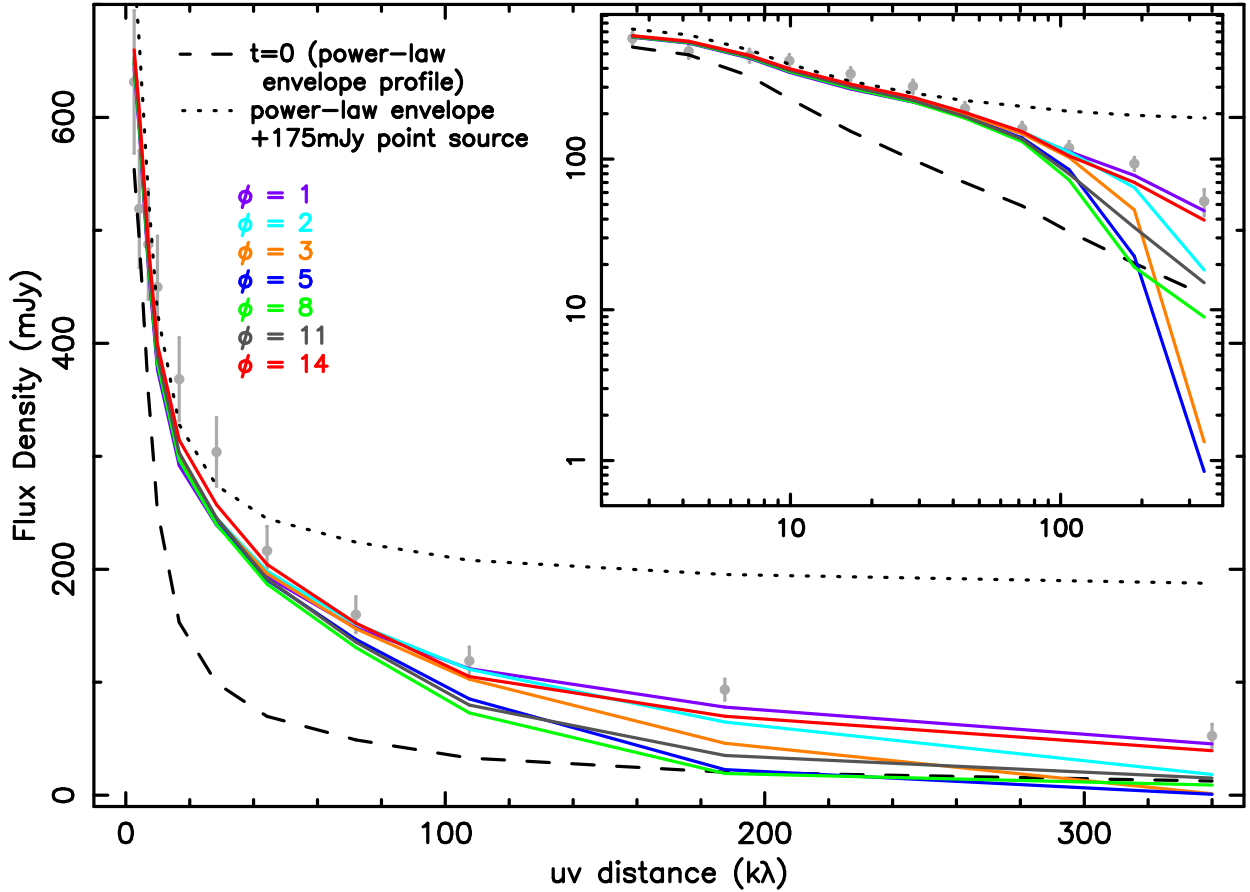}
\caption{
The solid curves show the example {\it u-v} visibilities 
from an optically thin envelope 
with outer radius 5000 AU, mass 5 $M_\odot$, and various evolutionary phases   
(no point source flux is added, and observational parameters such as the distance and 
{\it u-v} samplings are assumed to be the same as NGC 1333 IRAS 4A). 
Different colors correspond to different phases in the TM2005 model, 
as the density profiles of these phases are plotted in Fig. \ref{figDenProf}. 
The dashed curve is the flux density of the same case but with the initial 
density profile of TM2005, which is a nearly power-law profile with index -1.7. 
The dotted curve is the same nearly power-law envelope emission 
plus a constant 175 mJy (a point source flux) representing an unresolved disk. 
The same visibilities are also plotted in a logarithmic scale in the inset panel. 
The observational data points of NGC 1333 IRAS 4A are shown for comparison.  
The power-law envelope itself ({\it the dashed curve}) is not a good fit; 
also, just adding a point source flux 
to the power-law density profile cannot fit the data ({\it the dotted curve}). 
This is similar to the results from \citet{Jorgensen2005}.  
However, the predicted envelope emission from the TM2005 model is very different 
and a better fit to the data. 
}
\label{figDenUV}
\end{figure}

For each object, we did a parameter fitting of the model 
to the observational data with four degrees of freedom: 
evolutionary phase (density profiles) in the TM2005 magnetic cycle, 
outer cutoff radius of the envelope, 
central point source flux,  
and envelope mass. 
The inner cutoff radius is fixed to be 10 AU to be consistent with  
the central sink approximation used in TM2005 and 
also the inner envelope is truncated by the central disk physically. 
In Fig. \ref{figDenUV} we show the evolutionary phases in {\it u-v} 
space with all other parameters fixed for an example case of outer radius 
5000 AU, envelope mass 5 $M_\odot$, no point source flux, and a fixed 
power-law temperature profile of index -0.4 and 500 K at 1 AU as 
an optically thin case ({\it the solid curves}). 
The difference between the phases in {\it u-v} space is mainly at large  
{\it u-v} spacings, corresponding to the shock propagation in the 
density profiles.  
On the other hand, adding a point source is like adding a constant 
in {\it u-v} space, which effects the slope in logarithmic plots; 
increasing the envelope mass increases amplitude in all 
{\it u-v} spacings, especially in short {\it u-v} spacings; 
changing the outer radius alters the overall level and also 
the distribution of {\it u-v} amplitudes.  
The effects from varying each single parameter intertwine together and 
degeneracy makes it difficult to point out which parameter is the key to each good fit; moreover, 
the self-consistent temperature profile considers the increase of optical
depth near the center of the envelope and 
influences the predicted emission. 

Model images of the envelope are computed 
with consideration of the envelope emission and the
central point source attenuated by the dust envelope.
The self-consistent temperature profiles are calculated 
from the assumed luminosity, which was derived from the 
far-infrared flux density (see LMW2003) and listed in column 2 
in Table \ref{tabFit}.
The model images are multiplied by the observational primary beam, 
Fourier transformed into visibilities, 
and sampled with the same {\it u-v} coverage as the observations.  
Both model and observational data are binned and averaged in {\it u-v} annuli, 
shown as the flux density as a function of {\it u-v} spacing, and 
compared to each other. 
A reduced $\chi^2_r$ is computed to determine the goodness of a fit.  
Among the four degrees of freedom, the total envelope mass is adjusted 
to minimize $\chi^2_r$ while the other three parameters are fixed.  
This nonlinear minimization is done 
for any combination of these three parameters 
with outer radius ranging from 2000 to 9000 AU and point source flux 
as the ranges given in the parentheses in Table \ref{tabFit}, 
and for each set of parameters a total envelope mass was obtained
with local minimal $\chi^2_r$. 
Collecting the results of sets of parameters, the best fit, 
with global minimal $\chi^2_r$ is found.

\section{Results and Discussion}
\begin{figure}  
\includegraphics[angle=270,width=1.0\textwidth]{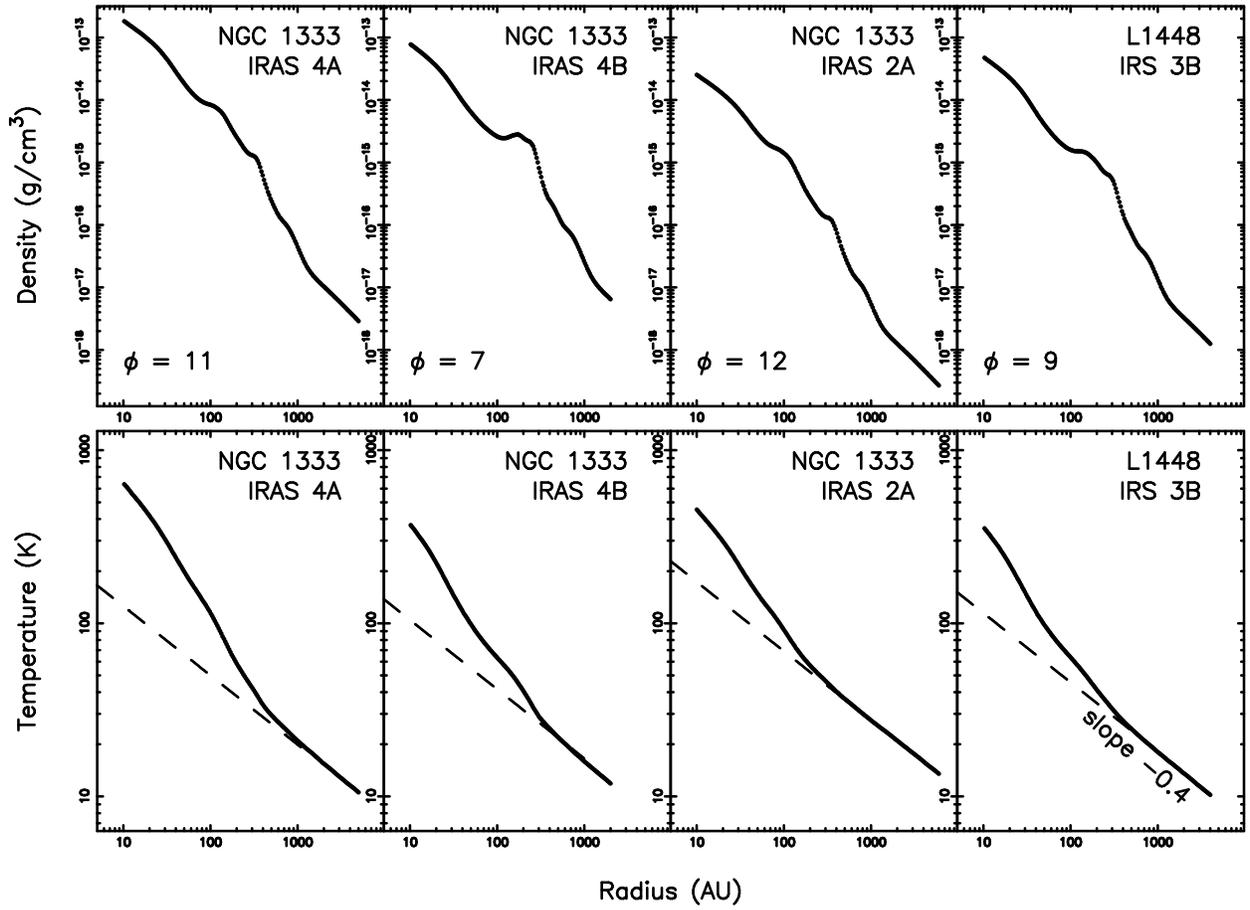}
\caption{
The density and temperature profiles of the best fit for each source. The  
straight dashed lines in the temperature plots show the optically thin case. 
}
\label{figDenTemp}
\end{figure}

In Fig. \ref{figfit}, data are binned, averaged, and shown by asterisks 
with associated error bars, and 
the curves show the best fit for each source   
as a function of u-v distance 
with the best fit parameters given on the plot. 
The best fit parameters do not need to be the same as those in LMW2003 
since a different theoretical model is used here. 
Fig. \ref{figDenTemp} shows the corresponding density and 
temperature profiles for these best fits. 
The straight dashed lines in the temperature plots are 
lines with slope $-0.4$, 
which would be the temperature profiles if the envelopes are optically thin
($T \propto r^{-0.4}$). 
We can see that the outer part of the envelopes is nearly optically thin, 
and becomes hotter and optically thicker at smaller radii. 

We refer to a fit 
with more than 90\% confidence level  
as an acceptable fit, and a summary of acceptable fits is given in 
Table \ref{tabFit}. 
Fitting with density profiles at different phases implies that  
the systems may be at different phases of a magnetic cycle.  
Presumably, if we had data of more objects we might catch systems at all phases.
The acceptable fits spread over a range of parameters depending on the 
signal-to-noise ratio of the observations.
One important aspect of the modeling is the clear need
for high signal-to-noise data.
The $\chi^2$ value is the evaluation quantity of 
the goodness of a set of parameters, 
and is smaller when the observational uncertainties are larger;  
low signal-to-noise ratio data make acceptable fits easier.   
When the signal-to-noise ratio is too low, 
the fitting becomes meaningless, which is why we 
apply the analysis only  
to four sources from LMW2000.
This argument is consistent with our results that 
the best constraints of model parameters is provided by 
the highest signal-to-noise observation IRAS 4A among all four sources.  
Given better signal-to-noise ratio data at more wavelengths
in the future 
we may be able to constrain the model better 
not only for more sources but also for better-constrained parameters.   

Nonetheless, the most important result of this study
is that with the TM2005 theoretical model,
we can easily fit the observations of these sources 
without conflicts of ages, unlike 
the fits using the LP or Shu models (LMW2003), although the
exact source age is not determinable by comparison with the TM2005 model.
Due to the episodic nature of the theoretical model, we are not fitting nor
implying a specific age for the source, but the range of age is 15 to 255 kyr. 
Parsec-scale jets and outflows have been found in these regions
and can be used to estimate the ages of central sources
that drive the outflows.
For example, the timescale of the outflows in NGC 1333 is of order $10^5$ yr  
\citep{Bally1996}, 
which is consistent with the age range of the TM2005 model.  

We use the phases of a typical magnetic cycle for the fitting.  
TM2005b Fig. 5 shows the beginning evolution after the central sink 
is introduced and the system has the first magnetic cycle 
at about {\it t}~=~15 kyr; then it 
repeats the magnetic cycles as in TM2005b Fig. 8 and 9
until {\it t}~=~255 kyr when the sink mass reaches
1 $M_{\odot}$ and the simulation stops.
Once the cycling behavior has been well established, all physical quantities
show high similarity of variation from cycle to cycle. 
Only the period of a cycle,
which is about several thousand years, 
decreases with time, controlled by the ambipolar-diffusion
timescale at the position of the magnetic wall; 
other than that, there is no obvious difference between cycles.  
So given a fit to a density profile, the system may be 
at a specific phase of any cycle during this epoch.
However, the age range suggested by the theoretical model is consistent with 
the age estimated by the outflows' scale, which implies that 
the observational data are consistent with
the TM2005 picture of early star formation.

One of the most important differences 
between the TM2005 and the Shu or LP models is 
that the steep power-law-like density profiles of TM2005 are actually not 
in equilibrium, while $\rho \propto r^{-2}$ in the other models corresponding
to the singular isothermal sphere is a critical equilibrium situation.  
In the Shu or LP models, the collapse is induced by an outward-moving 
rarefaction wave and the density  
changes dramatically after being affected by the 
wave; in TM2005, the system is collapsing and 
mass keeps accreting onto the central protostar dynamically 
without making an abrupt change,  
except for the relatively small bump generated by the magnetic wall.   
Periodic creation and dispersion of the magnetic wall dominate 
the variation of density profiles and also the accretion, so  
the whole collapse process is regulated by magnetic forces.  
Accounting for magnetic fields in the theoretical models, 
the density profiles at later times of 
the evolution are very different from those of the Shu or LP models, 
and are better matched with the observations and estimated ages from outflows. 

Can we constrain the earliest disks?
Table \ref{tabFit} shows that in all cases, we can fit our sources without
a circumstellar disk component at the 90\% confidence level.
On the other hand, a disk is expected early on in the collapse process
due to the rotation and/or magnetic fields in the initial cloud.
As material accretes, the
disk receives more mass from the envelope.  The mass is processed in the disk,
which regulates mass flow through the disk and onto the protostar.
The magnetic fields within the disk and star
give rise to an outflow \citep[e.g.,][]{shang2007,Pudritz2007PPV}
that is typically seen even in very young sources,
suggesting that a disk is established nearly concurrently with the
protostar's growth in luminosity as it collapses to stellar size.
In fact, simple theoretical arguments suggest that the disk
evolves early and grows quickly with time as
$\sim t^2$ or $\sim t^3$ \citep[e.g.,][]{stahlerbook}, which depends crucially 
on the mass accretion rate.

Although not a statistical requirement, we do expect these sources
to have some sort of deeply embedded circumstellar disks in the center of the infalling envelopes.
We can use HL Tau, which was modeled in \citet{Mundy1996}, 
as a standard candle to estimate the mass of the disks in our modeled systems.  
Also using BIMA $\lambda$~=~2.7~mm dust continuum observations, 
the total flux from HL Tau was $\sim$100 mJy, 
and the derived disk mass from modeling was $\sim$0.05 $M_{\odot}$.  
Based on the assumption of equal flux to disk mass ratio corrected by the 
distance, the value 0.010 Jy in the best fit point source flux for 
NGC 1333 IRAS 4A corresponds to a disk of mass 0.03 $M_{\odot}$, 
and 0.035 Jy for IRAS 4B corresponds to a 0.11 $M_{\odot}$ disk.  
A distance to NGC 1333 of 350 pc has been assumed here. 
If we use a distance of 250 pc instead, 
the estimated mass of embedded disks for the best fits 
would be 0.016 $M_{\odot}$ and 0.056 $M_{\odot}$
for IRAS 4A and IRAS 4B, respectively.  
The acceptable fits with maximum disk components give disk mass of 
0.16 $M_{\odot}$, 
0.18 $M_{\odot}$, 
0.065 $M_{\odot}$, and  
0.041 $M_{\odot}$ 
for NGC 1333 IRAS 4A, IRAS 4B, IRAS 2A, and L1448 IRS 3B, respectively.  
It gives an upper limit of disk mass to IRAS 2A 
since model parameters with higher point source flux cannot fit anymore.  

Unlike \citet{Jorgensen2005},  
a disk component is not crucially necessary in our modeling. 
The major reason is that different envelope models are used.  
TM2005 predict a very different 
visibility amplitude profile from what is expected by a power-law density profile. 
For example, in Fig. \ref{figDenUV}, the dashed curve 
is the predicted {\it u-v} amplitude of a nearly power-law density profile of 
index -1.7 (the initial density profile in the TM2005 simulation), and  
the dotted curve is generated with the same density profile, 
but with a point source flux representing an unresolved circumstellar disk.  
As can be seen, the envelope emission from a power-law density profile shows a very different shape 
than those predicted by the phases of the magnetic cycle.   
TM2005 density profiles are able to fit the data well 
without adding a Gaussian disk \citep[cf.][]{Jorgensen2005}. 
Again this shows the importance of understanding the collapsing envelope in order to 
understand the embedded disk.

\section{Summary}
\begin{enumerate} 
\item
Although interferometry is a powerful tool in
resolving out the large-scale emission of the envelope, 
the ability to detect a disk in the presence
of an envelope is not set by the formal noise level or Fourier components, 
but by the intrinsic ability to model the complexity
of the envelope (inner cutoff, asymmetric structure, etc.)
and by the resolution of the observations compared to the disk
size and the inner cutoff of the envelope.
In other words, reliable detection of the embedded circumstellar disk
requires a knowledge of the physical parameters of the inner envelope.

\item 
Our observational data are consistent with the theoretical predictions of 
TM2005 concerning the density profiles.  Moreover, 
there is no discrepancy in age based on the size of the outflows, unlike  
the fitting results of 
the simpler Shu or LP solutions to the isothermal sphere (e.g., LMW2003). 
It is important to note that the exact age cannot be determined 
by comparison of the data with the TM2005 model due to its periodic nature.  
Regardless of the collapse initiation, the
magnetic fields may play an important role early on in the collapse process.
To expand this comparison, we will continue to observe Class 0 objects at higher
sensitivity and multiple wavelengths, better incorporating the theoretical models into our
comparisons.

\item 
Although our initial results do not require the existence of circumstellar
disks (acceptable fits of 0.0 to 0.11 $M_\odot$), we can place upper
limits on the disk masses. In general, the disks are less massive than
$\sim$0.1 $M_{\odot}$. The youngest circumstellar disk mass is not overly
massive compared to other well known Class II or III circumstellar disks.
The fact that a disk component is not statistically necessary in this
modeling is different from similar work at sub-millimeter wavelengths by
\citet{Jorgensen2005}. The main reason is that different visibility
profiles are predicted by TM2005 than for envelopes of power-law density.

\end{enumerate} 

\acknowledgements{
HFC and LWL acknowledge support from the Laboratory for
Astronomical Imaging at the University of Illinois and
the NSF under Grant No. AST-05-40459 and NASA Origins Grant No. NNG06GE41G.
KT acknowledges support by NSF Grants AST-02-06216 and 
AST-02-39759,  by the NASA Theoretical Astrophysics Program Grant
NNG04G178G and  by the Kavli Institute for Cosmological Physics at the
University of Chicago through NSF Grants PHY-01-14422 and PHY-05-51142 and 
an endowment from the Kavli Foundation and its founder Fred Kavli.
LGM acknowledges NSF Grant No. AST-05-40450 and NASA Origins Grant
No. NNG06GE16G.
HFC, LWL, and TM acknowledge support from the NSF under Grant 
No. AST-07-09206. 
Helpful comments by an anonymous referee are gratefully acknowledged.
}

\bibliographystyle{apj}
\bibliography{ref}

\begin{thebibliography}{64}
\expandafter\ifx\csname natexlab\endcsname\relax\def\natexlab#1{#1}\fi

\bibitem[{{Adams} {et~al.}(1987){Adams}, {Lada}, \& {Shu}}]{Adams1987}
{Adams}, F.~C., {Lada}, C.~J., \& {Shu}, F.~H. 1987, \apj, 312, 788

\bibitem[{{Andre} {et~al.}(1993){Andre}, {Ward-Thompson}, \&
  {Barsony}}]{Andre1993}
{Andre}, P., {Ward-Thompson}, D., \& {Barsony}, M. 1993, \apj, 406, 122

\bibitem[{{Andre} {et~al.}(2000){Andre}, {Ward-Thompson}, \&
  {Barsony}}]{Andre2000PPIV}
{Andre}, P., {Ward-Thompson}, D., \& {Barsony}, M. 2000, in Protostars and
  Planets IV, ed. V.~{Mannings}, A.~{Boss}, \& S.~S. {Russell}, 59

\bibitem[{{Arce} \& {Sargent}(2006)}]{Arce2006}
{Arce}, H.~G. \& {Sargent}, A.~I. 2006, \apj, 646, 1070

\bibitem[{{Bachiller}(1996)}]{Bachiller1996ar}
{Bachiller}, R. 1996, \araa, 34, 111

\bibitem[{{Bally} {et~al.}(1996){Bally}, {Devine}, \& {Reipurth}}]{Bally1996}
{Bally}, J., {Devine}, D., \& {Reipurth}, B. 1996, \apjl, 473, L49

\bibitem[{{Barranco} \& {Goodman}(1998)}]{bg98}
{Barranco}, J.~A. \& {Goodman}, A.~A. 1998, \apj, 504, 207

\bibitem[{{Beckwith} \& {Sargent}(1991)}]{Beckwith1991}
{Beckwith}, S.~V.~W. \& {Sargent}, A.~I. 1991, \apj, 381, 250

\bibitem[{{Benson} \& {Myers}(1989)}]{bm89}
{Benson}, P.~J. \& {Myers}, P.~C. 1989, \apjs, 71, 89

\bibitem[{{Blake} {et~al.}(1995){Blake}, {Sandell}, {van Dishoeck},
  {Groesbeck}, {Mundy}, \& {Aspin}}]{Blake1995}
{Blake}, G.~A., {Sandell}, G., {van Dishoeck}, E.~F., {Groesbeck}, T.~D.,
  {Mundy}, L.~G., \& {Aspin}, C. 1995, \apj, 441, 689

\bibitem[{{Brinch} {et~al.}(2007){Brinch}, {Crapsi}, {J{\o}rgensen},
  {Hogerheijde}, \& {Hill}}]{Brinch2007}
{Brinch}, C., {Crapsi}, A., {J{\o}rgensen}, J.~K., {Hogerheijde}, M.~R., \&
  {Hill}, T. 2007, \aap, 475, 915

\bibitem[{{Cernis}(1990)}]{Cernis1990}
{Cernis}, K. 1990, \apss, 166, 315

\bibitem[{{Cernis}(1993)}]{Cernis1993}
---. 1993, Baltic Astronomy, 2, 214

\bibitem[{{Choi}(2001)}]{Choi2001}
{Choi}, M. 2001, \apj, 553, 219

\bibitem[{{Choi}(2005)}]{Choi2005}
---. 2005, \apj, 630, 976

\bibitem[{{de Zeeuw} {et~al.}(1999){de Zeeuw}, {Hoogerwerf}, {de Bruijne},
  {Brown}, \& {Blaauw}}]{deZeeuw1999}
{de Zeeuw}, P.~T., {Hoogerwerf}, R., {de Bruijne}, J.~H.~J., {Brown}, A.~G.~A.,
  \& {Blaauw}, A. 1999, \aj, 117, 354

\bibitem[{{Desch} \& {Mouschovias}(2001)}]{Desch2001}
{Desch}, S.~J. \& {Mouschovias}, T.~{\relax Ch}. 2001, \apj, 550, 314

\bibitem[{{Di Francesco} {et~al.}(2001){Di Francesco}, {Myers}, {Wilner},
  {Ohashi}, \& {Mardones}}]{DiFrancesco2001}
{Di Francesco}, J., {Myers}, P.~C., {Wilner}, D.~J., {Ohashi}, N., \&
  {Mardones}, D. 2001, \apj, 562, 770

\bibitem[{{Draine} \& {Lee}(1984)}]{Draine1984}
{Draine}, B.~T. \& {Lee}, H.~M. 1984, \apj, 285, 89

\bibitem[{{Elmegreen} \& {Scalo}(2004)}]{Elmegreen2004}
{Elmegreen}, B.~G. \& {Scalo}, J. 2004, \araa, 42, 211

\bibitem[{{Enoch} {et~al.}(2006){Enoch}, {Young}, {Glenn}, {Evans}, {Golwala},
  {Sargent}, {Harvey}, {Aguirre}, {Goldin}, {Haig}, {Huard}, {Lange},
  {Laurent}, {Maloney}, {Mauskopf}, {Rossinot}, \& {Sayers}}]{Enoch2006}
{Enoch}, M.~L., {Young}, K.~E., {Glenn}, J., {Evans}, II, N.~J., {Golwala}, S.,
  {Sargent}, A.~I., {Harvey}, P., {Aguirre}, J., {Goldin}, A., {Haig}, D.,
  {Huard}, T.~L., {Lange}, A., {Laurent}, G., {Maloney}, P., {Mauskopf}, P.,
  {Rossinot}, P., \& {Sayers}, J. 2006, \apj, 638, 293

\bibitem[{{Fiedler} \& {Mouschovias}(1993)}]{fiedler1993}
{Fiedler}, R.~A. \& {Mouschovias}, T.~{\relax Ch}. 1993, \apj, 415, 680

\bibitem[{{Girart} {et~al.}(2006){Girart}, {Rao}, \& {Marrone}}]{Girart2006}
{Girart}, J.~M., {Rao}, R., \& {Marrone}, D.~P. 2006, Science, 313, 812

\bibitem[{{Harvey} {et~al.}(2003){Harvey}, {Wilner}, {Myers}, {Tafalla}, \&
  {Mardones}}]{Harvey2003a}
{Harvey}, D.~W.~A., {Wilner}, D.~J., {Myers}, P.~C., {Tafalla}, M., \&
  {Mardones}, D. 2003, \apj, 583, 809

\bibitem[{{Hunter}(1977)}]{Hunter1977}
{Hunter}, C. 1977, \apj, 218, 834

\bibitem[{{J{\o}rgensen} {et~al.}(2005){J{\o}rgensen}, {Bourke}, {Myers},
  {Sch{\"o}ier}, {van Dishoeck}, \& {Wilner}}]{Jorgensen2005}
{J{\o}rgensen}, J.~K., {Bourke}, T.~L., {Myers}, P.~C., {Sch{\"o}ier}, F.~L.,
  {van Dishoeck}, E.~F., \& {Wilner}, D.~J. 2005, \apj, 632, 973

\bibitem[{{J{\o}rgensen} {et~al.}(2006){J{\o}rgensen}, {Harvey}, {Evans},
  {Huard}, {Allen}, {Porras}, {Blake}, {Bourke}, {Chapman}, {Cieza}, {Koerner},
  {Lai}, {Mundy}, {Myers}, {Padgett}, {Rebull}, {Sargent}, {Spiesman},
  {Stapelfeldt}, {van Dishoeck}, {Wahhaj}, \& {Young}}]{Jorgensen2006}
{J{\o}rgensen}, J.~K., {Harvey}, P.~M., {Evans}, II, N.~J., {Huard}, T.~L.,
  {Allen}, L.~E., {Porras}, A., {Blake}, G.~A., {Bourke}, T.~L., {Chapman}, N.,
  {Cieza}, L., {Koerner}, D.~W., {Lai}, S.-P., {Mundy}, L.~G., {Myers}, P.~C.,
  {Padgett}, D.~L., {Rebull}, L., {Sargent}, A.~I., {Spiesman}, W.,
  {Stapelfeldt}, K.~R., {van Dishoeck}, E.~F., {Wahhaj}, Z., \& {Young}, K.~E.
  2006, \apj, 645, 1246

\bibitem[{{J{\o}rgensen} {et~al.}(2004){J{\o}rgensen}, {Hogerheijde}, {van
  Dishoeck}, {Blake}, \& {Sch{\"o}ier}}]{Jorgensen2004_IRAS2}
{J{\o}rgensen}, J.~K., {Hogerheijde}, M.~R., {van Dishoeck}, E.~F., {Blake},
  G.~A., \& {Sch{\"o}ier}, F.~L. 2004, \aap, 413, 993

\bibitem[{{Keene} \& {Masson}(1990)}]{Keene1990}
{Keene}, J. \& {Masson}, C.~R. 1990, \apj, 355, 635

\bibitem[{{Kirk} {et~al.}(2007){Kirk}, {Johnstone}, \& {Tafalla}}]{kjt07}
{Kirk}, H., {Johnstone}, D., \& {Tafalla}, M. 2007, ArXiv e-prints, 707

\bibitem[{{Knee} \& {Sandell}(2000)}]{Knee2000}
{Knee}, L.~B.~G. \& {Sandell}, G. 2000, \aap, 361, 671

\bibitem[{{Kwon} {et~al.}(2006){Kwon}, {Looney}, {Crutcher}, \&
  {Kirk}}]{Kwon2006}
{Kwon}, W., {Looney}, L.~W., {Crutcher}, R.~M., \& {Kirk}, J.~M. 2006, \apj,
  653, 1358

\bibitem[{{Lada} {et~al.}(1996){Lada}, {Alves}, \& {Lada}}]{Lada1996}
{Lada}, C.~J., {Alves}, J., \& {Lada}, E.~A. 1996, \aj, 111, 1964

\bibitem[{{Lada} \& {Wilking}(1984)}]{Lada1984}
{Lada}, C.~J. \& {Wilking}, B.~A. 1984, \apj, 287, 610

\bibitem[{{Larson}(1969)}]{Larson1969}
{Larson}, R.~B. 1969, \mnras, 145, 271

\bibitem[{{Lay} {et~al.}(1995){Lay}, {Carlstrom}, \& {Hills}}]{Lay1995}
{Lay}, O.~P., {Carlstrom}, J.~E., \& {Hills}, R.~E. 1995, \apjl, 452, L73

\bibitem[{{Looney} {et~al.}(2000){Looney}, {Mundy}, \& {Welch}}]{LMW2000}
{Looney}, L.~W., {Mundy}, L.~G., \& {Welch}, W.~J. 2000, \apj, 529, 477

\bibitem[{{Looney} {et~al.}(2003){Looney}, {Mundy}, \& {Welch}}]{LMW2003}
---. 2003, \apj, 592, 255

\bibitem[{{Looney} {et~al.}(2007){Looney}, {Tobin}, \& {Kwon}}]{looney2007}
{Looney}, L.~W., {Tobin}, J.~J., \& {Kwon}, W. 2007, \apjl, 670, L131

\bibitem[{{Mac Low} \& {Klessen}(2004)}]{MacLow2004}
{Mac Low}, M.-M. \& {Klessen}, R.~S. 2004, Reviews of Modern Physics, 76, 125

\bibitem[{{Mathis} {et~al.}(1977){Mathis}, {Rumpl}, \& {Nordsieck}}]{MRN1977}
{Mathis}, J.~S., {Rumpl}, W., \& {Nordsieck}, K.~H. 1977, \apj, 217, 425

\bibitem[{{Mouschovias}(1996)}]{Mouschovias1996}
{Mouschovias}, T.~{\relax Ch}. 1996, in The Role of Dust in the Formation of
  Stars, ed. H.~U. {K{\"a}ufl} \& R.~{Siebenmorgen} (Berlin: Springer-Verlag),
  382

\bibitem[{{Mouschovias} \& {Ciolek}(1999)}]{Mouschovias1999}
{Mouschovias}, T.~{\relax Ch}. \& {Ciolek}, G.~E. 1999, in The Origin of Stars
  and Planetary Systems, ed. C.~J. {Lada} \& N.~D. {Kylafis} (Dordrecht: Kluwer
  Academic Publishers), 305

\bibitem[{{Mundy} {et~al.}(1996){Mundy}, {Looney}, {Erickson}, {Grossman},
  {Welch}, {Forster}, {Wright}, {Plambeck}, {Lugten}, \&
  {Thornton}}]{Mundy1996}
{Mundy}, L.~G., {Looney}, L.~W., {Erickson}, W., {Grossman}, A., {Welch},
  W.~J., {Forster}, J.~R., {Wright}, M.~C.~H., {Plambeck}, R.~L., {Lugten}, J.,
  \& {Thornton}, D.~D. 1996, \apjl, 464, L169

\bibitem[{{Natta} {et~al.}(2007){Natta}, {Testi}, {Calvet}, {Henning},
  {Waters}, \& {Wilner}}]{Natta2007PPV}
{Natta}, A., {Testi}, L., {Calvet}, N., {Henning}, T., {Waters}, R., \&
  {Wilner}, D. 2007, in Protostars and Planets V, ed. B.~{Reipurth},
  D.~{Jewitt}, \& K.~{Keil}, 767--781

\bibitem[{{O'Linger} {et~al.}(2006){O'Linger}, {Cole}, {Ressler}, \&
  {Wolf-Chase}}]{OLinger2006}
{O'Linger}, J.~C., {Cole}, D.~M., {Ressler}, M.~E., \& {Wolf-Chase}, G. 2006,
  \aj, 131, 2601

\bibitem[{{Penston}(1969)}]{Penston1969}
{Penston}, M.~V. 1969, \mnras, 144, 425

\bibitem[{{Pudritz} {et~al.}(2007){Pudritz}, {Ouyed}, {Fendt}, \&
  {Brandenburg}}]{Pudritz2007PPV}
{Pudritz}, R.~E., {Ouyed}, R., {Fendt}, C., \& {Brandenburg}, A. 2007, in
  Protostars and Planets V, ed. B.~{Reipurth}, D.~{Jewitt}, \& K.~{Keil},
  277--294

\bibitem[{{Sandell} \& {Knee}(2001)}]{Sandell2001}
{Sandell}, G. \& {Knee}, L.~B.~G. 2001, \apjl, 546, L49

\bibitem[{{Seale} \& {Looney}(2007)}]{Seale2008}
{Seale}, J.~P. \& {Looney}, L.~W. 2007, ArXiv e-prints, 711

\bibitem[{{Shang} {et~al.}(2007){Shang}, {Li}, \& {Hirano}}]{shang2007}
{Shang}, H., {Li}, Z.-Y., \& {Hirano}, N. 2007, in Protostars and Planets V,
  ed. B.~{Reipurth}, D.~{Jewitt}, \& K.~{Keil}, 261--276

\bibitem[{{Shu}(1977)}]{Shu1977}
{Shu}, F.~H. 1977, \apj, 214, 488

\bibitem[{{Stahler} \& {Palla}(2005)}]{stahlerbook}
{Stahler}, S.~W. \& {Palla}, F. 2005, {The Formation of Stars} (Weinheim:
  Wiley-VCH)

\bibitem[{{Tassis} \& {Mouschovias}(2005{\natexlab{a}})}]{Tassis2005a}
{Tassis}, K. \& {Mouschovias}, T.~{\relax Ch}. 2005{\natexlab{a}}, \apj, 618,
  769

\bibitem[{{Tassis} \& {Mouschovias}(2005{\natexlab{b}})}]{Tassis2005b}
---. 2005{\natexlab{b}}, \apj, 618, 783

\bibitem[{{Tassis} \& {Mouschovias}(2007{\natexlab{a}})}]{tm07a}
---. 2007{\natexlab{a}}, \apj, 660, 370

\bibitem[{{Tassis} \& {Mouschovias}(2007{\natexlab{b}})}]{tm07b}
---. 2007{\natexlab{b}}, \apj, 660, 388

\bibitem[{{Tassis} \& {Mouschovias}(2007{\natexlab{c}})}]{tm07c}
---. 2007{\natexlab{c}}, \apj, 660, 402

\bibitem[{{Tobin} {et~al.}(2007){Tobin}, {Looney}, {Mundy}, {Kwon}, \&
  {Hamidouche}}]{Tobin2007}
{Tobin}, J.~J., {Looney}, L.~W., {Mundy}, L.~G., {Kwon}, W., \& {Hamidouche},
  M. 2007, \apj, 659, 1404

\bibitem[{{Whitworth} \& {Summers}(1985)}]{Whitworth1985}
{Whitworth}, A. \& {Summers}, D. 1985, \mnras, 214, 1

\bibitem[{{Wilking} {et~al.}(2004){Wilking}, {Meyer}, {Greene}, {Mikhail}, \&
  {Carlson}}]{Wilking2004}
{Wilking}, B.~A., {Meyer}, M.~R., {Greene}, T.~P., {Mikhail}, A., \& {Carlson},
  G. 2004, \aj, 127, 1131

\bibitem[{{Wilner} \& {Welch}(1994)}]{Wilner1994}
{Wilner}, D.~J. \& {Welch}, W.~J. 1994, \apj, 427, 898

\bibitem[{{Wolf-Chase} {et~al.}(2000){Wolf-Chase}, {Barsony}, \&
  {O'Linger}}]{Wolf-Chase2000}
{Wolf-Chase}, G.~A., {Barsony}, M., \& {O'Linger}, J. 2000, \aj, 120, 1467

\bibitem[{{Wolfire} \& {Cassinelli}(1986)}]{WC1986}
{Wolfire}, M.~G. \& {Cassinelli}, J.~P. 1986, \apj, 310, 207

\end{thebibliography}

\begin{deluxetable}{ccccccc} 
\tabletypesize{\scriptsize}
\tablecolumns{6} 
\tablewidth{0pc} 
\tablecaption{Acceptable Fits} 
\tablehead{ 
\colhead{Source}         & 
\colhead{Luminosity }    & 
\colhead{Density Profiles}    & \colhead{Outer Radius} & 
\colhead{Point Source}         & \colhead{Envelope Mass}   & 
\colhead{Good Fits \tablenotemark{d}  }
\\
\colhead{Name}         &
\colhead{(L$_\odot$) \tablenotemark{a} }         &
\colhead{(Phase) \tablenotemark{b} }    & \colhead{(AU)} &
\colhead{Flux (mJy) \tablenotemark{c}}     & \colhead{(M$_\odot$)}   &
\colhead{}   
}
\startdata 
NGC 1333 IRAS 4A & 16  &  9-11 & 4000-5000   & 
 0-50 (0-50)   & 4.88-6.23 & 
15/720 \\     
NGC 1333 IRAS 4B & 5.2  &  1-14 & 2000-6000   & 
 0-56 (0-56)  & 1.45-4.02  & 
302/1215 \\   
NGC 1333 IRAS 2A  & 30  &  2-13 & 5000-8000   & 
 0-7  (0-21) & 0.76-1.37 & 
49/480 \\   
L1448 IRS 3B & 6.8 &  3-12 & 2000-9000  & 
 0-18  (0-18) & 0.75-4.37 & 
223/480 \\   
\enddata 
\tablecomments{
A set of parameters gives a good fit (a so-called acceptable fit) 
if the reduced $\chi^2$ is within the 90\% confidence level.  
For each parameter the range of acceptable fits is given.
}
\tablenotetext{a}{
All sources here are binary system and flux of single component is 
assumed based on the ratio of fluxes at $\lambda$~=~2.7~mm (see LMW2003). 
}
\tablenotetext{b}{
The number corresponds to a specific phase in the cycle for which the 
density profile of the TM2005 model provides an acceptable fit 
(see Fig. \ref{figDenProf}).  
}
\tablenotetext{c}{
The parameter search ranges fitting are listed in the parentheses. 
}
\tablenotetext{d}{
The numbers of good fits and the total number of combinations while fitting.  
}
\label{tabFit}
\end{deluxetable} 

\end{document}